\documentclass[aps,amssymb,amsmath,pra,reprint,noshowpacs,superscriptaddress]{revtex4-2}
\usepackage[utf8]{inputenc}
\usepackage{amsthm}
\usepackage{mathtools}
\usepackage{xcolor}
\usepackage{graphicx}
\usepackage[left=23mm,right=13mm,top=35mm,columnsep=15pt]{geometry} 
\usepackage{adjustbox}
\usepackage{placeins}
\usepackage[T1]{fontenc}
\usepackage{lipsum}
\usepackage{csquotes}
\usepackage{siunitx}
\DeclareSIUnit\bar{bar}
\usepackage[english]{babel}
\usepackage{helvet}
\usepackage{scalefnt} 
\usepackage{natbib}
\usepackage{hyperref}
\usepackage{appendix}
\usepackage{color}
\usepackage{comment}
\hypersetup{
	colorlinks   = true, 
	urlcolor     = blue, 
	linkcolor    = blue,
	citecolor   = blue 
}
\usepackage{subfigure}
\begin{document}
\scalefont{1.0}

\title{Simultaneous ground-state cooling of two mechanical modes of a levitated nanoparticle}

\author{Johannes~\surname{Piotrowski} }
\email[equal contribution]{ }
\affiliation{Photonics Laboratory, ETH Z{\"u}rich, 8093 Z\"urich, Switzerland}
\author{Dominik~\surname{Windey}}
\email[equal contribution]{ }
\affiliation{Photonics Laboratory, ETH Z{\"u}rich, 8093 Z\"urich, Switzerland}
\author{Jayadev~\surname{Vijayan}}
\affiliation{Photonics Laboratory, ETH Z{\"u}rich, 8093 Z\"urich, Switzerland}
\author{Carlos~\surname{Gonzalez-Ballestero}}
\affiliation{Institute for Quantum Optics and Quantum Information, Austrian Academy of Sciences, A-6020 Innsbruck, Austria}
\affiliation{Institute for Theoretical Physics, University of Innsbruck, A-6020 Innsbruck, Austria}
\author{Andr\'es~\surname{de~los~R\'ios Sommer}}
\affiliation{Nanophotonic Systems Laboratory, ETH Z\"urich, 8092 Z\"urich, Switzerland}
\author{Nadine~\surname{Meyer}}
\affiliation{Nanophotonic Systems Laboratory, ETH Z\"urich, 8092 Z\"urich, Switzerland}
\author{Romain~\surname{Quidant}}
\affiliation{Nanophotonic Systems Laboratory, ETH Z\"urich, 8092 Z\"urich, Switzerland}
\affiliation{Quantum Center, ETH Z\"urich, 8083 Z\"urich, Switzerland}
\author{Oriol~\surname{Romero-Isart}}
\affiliation{Institute for Quantum Optics and Quantum Information, Austrian Academy of Sciences, A-6020 Innsbruck, Austria}
\affiliation{Institute for Theoretical Physics, University of Innsbruck, A-6020 Innsbruck, Austria}
\author{Ren\'e~\surname{Reimann}}
\affiliation{Quantum Research Centre, Technology Innovation Institute, Abu Dhabi, UAE}
\author{Lukas~\surname{Novotny}}
\affiliation{Photonics Laboratory, ETH Z{\"u}rich, 8093 Z\"urich, Switzerland}

\definecolor{darkred}{rgb}{0.6, 0.1, 0.1}
\newcommand{\jpcomment}[1]{{\color{darkred}{ *** JP: #1 ***}}}
\newcommand{\rrcomment}[1]{{\color{blue}{ *** RR: #1 ***}}}
\definecolor{darkgreen}{rgb}{0.1, 0.6, 0.1}
\newcommand{\ascomment}[1]{{\color{darkgreen}{ *** AS: #1 ***}}}

%\date{\today} % Leave empty to omit a date

\begin{abstract}
The quantum ground state of a massive mechanical system is a stepping stone for investigating macroscopic quantum states and building high fidelity sensors.
With the recent achievement of ground-state cooling of a single motional mode~\cite{Delic2020,Tebbenjohanns2021,Magrini2021,Kamba2022,Ranfagni2022}, levitated nanoparticles have entered the quantum domain~\cite{Gonzalez-Ballestero2021}. 
To overcome detrimental cross-coupling and decoherence effects, quantum control needs to be expanded to more system dimensions, but the effect of a decoupled dark mode has thus far hindered cavity-based ground state cooling of multiple mechanical modes~\cite{Genes2008, Toros2021}.
Here, we demonstrate two-dimensional (2D) ground-state cooling of an optically levitated nanoparticle. Utilising coherent scattering~\cite{Vuletic2000, Hosseini2017, Vuletic2001, Windey2019, Delic2019} into an optical cavity mode, we reduce the occupation numbers of two separate centre-of-mass modes to 0.83 and 0.81, respectively. 
By controlling the frequency separation and the cavity coupling strengths of the nanoparticle's mechanical modes, we show the transition from 1D to 2D ground-state cooling while avoiding the effect of dark modes.
Our results lay the foundations for generating quantum-limited high orbital angular momentum states %(``quantum orbits'')
with applications in rotation sensing.
The demonstrated 2D control, combined with already shown capabilities of ground-state cooling along the third motional axis~\cite{Tebbenjohanns2021,Magrini2021}, opens the door for full 3D ground-state cooling of a massive object.
\end{abstract}

\maketitle
\centerline{\textbf{Introduction}}

Testing the limits of quantum mechanics as the system size approaches macroscopic scales is one of the grand fundamental and engineering challenges in modern physics~\cite{Bose1999, Leggett2002, Marshall2003, Schlosshauer2008}. Levitated systems are ideal testbeds for exploring macroscopic quantum physics due to dynamical and full control over their trapping potential~\cite{Romero-Isart2011, Romero-Isart2011a, Neumeier2022, Weiss2021}. The motional ground state is the stepping stone for the preparation of quantum states that are delocalised over scales larger than the zero-point motion. 
Recently, the ground state of the centre-of-mass (COM) motion of a levitated nanoparticle has been reached along a single direction, using both passive feedback via an optical cavity~\cite{Delic2020, Ranfagni2022} as well as active measurement-based feedback~\cite{Tebbenjohanns2021, Magrini2021,Kamba2022}.

Even though ground-state cooling is typically necessary for preparing macroscopic quantum states, it is not sufficient, as these states are susceptible to decoherence.
Cross-coupling between a hot and the ground-state cooled COM mode provides a decoherence channel for the ground-state cooled target mode.
Cooling of the hot mode would mitigate this cross-coupling decoherence.
Additionally, a second ground-state cooled mode (ancilla mode) would be a powerful tool to understand, or even compensate, decoherence effects which influence both ground-state cooled modes.
As an example, common dephasing sources (e.g. trap frequency noise) could be measured in the ancilla mode and then be counteracted in the target mode.

In our levitated particle setup we achieve simultaneous ground-state cooling of two mechanical modes. We circumvent the key obstacle in optomechanical systems preventing multimode ground-state cooling~\cite{Cattiaux2021, Liu2022}, namely the formation of dark modes~\cite{Genes2008, Shkarin2014, Ockeloen-Korppi2019,Lai2020}.
%Our levitated particle setup allows us to achieve simultaneous ground-state cooling of two mechanical modes. in an optomechanical system~\cite{Cattiaux2021, Liu2022}. 
%We circumvent the key obstacle in clamped optomechanical systems, namely the formation of dark modes~\cite{Genes2008, Shkarin2014, Ockeloen-Korppi2019}.
To this end we use the unique tunability of levitated systems and adjust the frequency difference between the involved mechanical modes to be larger than the optomechanical coupling rate~\cite{Genes2008, Toros2021}.
Under this constraint, we optimise the optomechanical coupling strength for efficient cooling by coherent scattering~\cite{Windey2019, Delic2019}.
\\

\centerline{\textbf{Experimental setup}}

\begin{figure}
    \centering
    \includegraphics[]{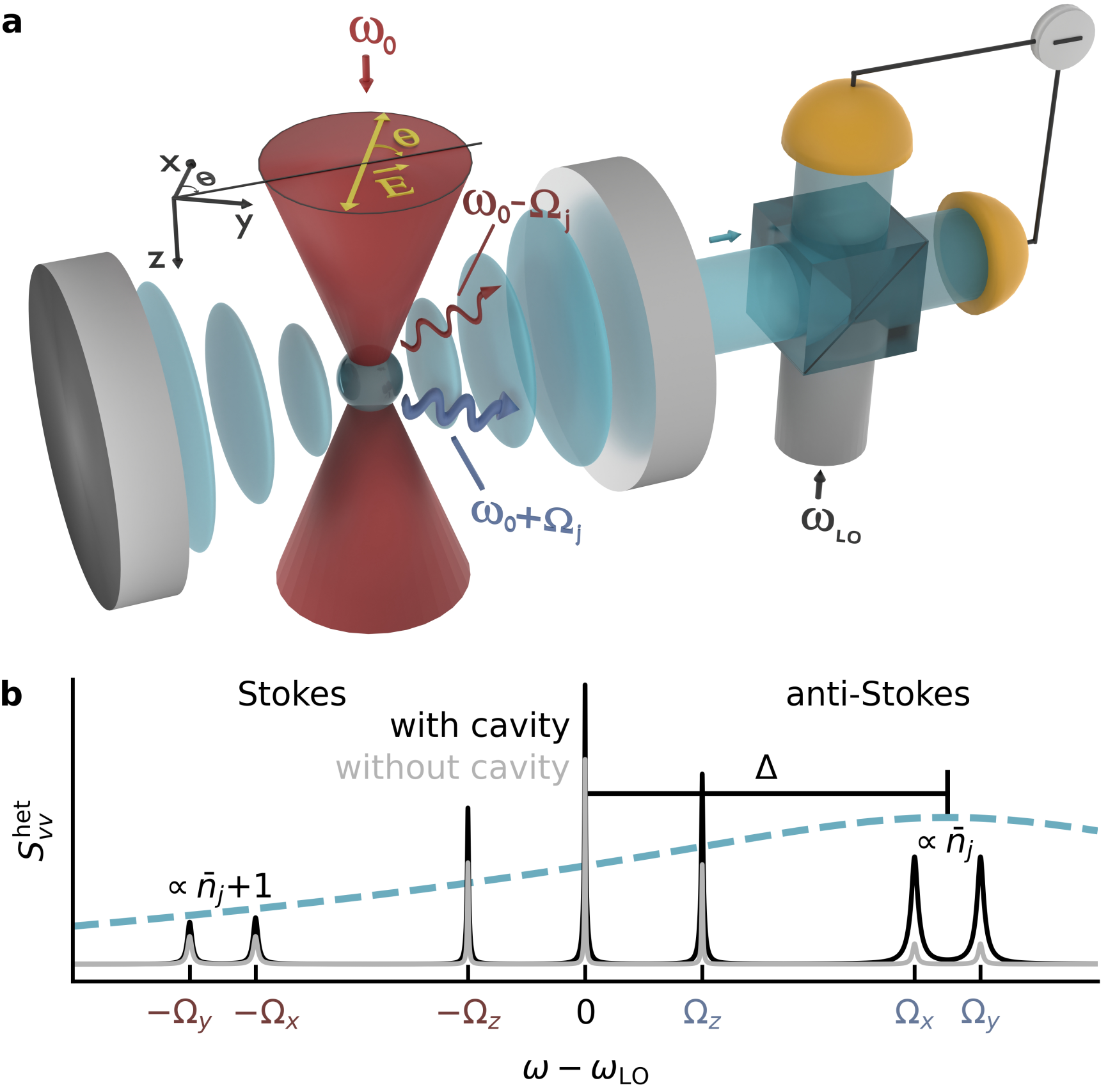}
    \caption{\textbf{Cooling and detection by coherent scattering.}
    \textbf{a},~Schematic of a cavity-coupled levitated nanoparticle. 
    The cavity enhances anti-Stokes (blue arrow) over Stokes (red arrow) scattering, leading to cooling of the nanoparticle's mechanical modes.
    The scattered light at $\omega_0\pm \Omega_{x,y,z}$ leaks through the high transmission mirror, and interferes with a strong local oscillator at $\omega_\mathrm{LO}$ in a heterodyne detection scheme. 
    The optical tweezers' propagation direction and its polarisation define the $z$ and $x$ axes, respectively. %, with $y$ being perpendicular to both. 
    The polarisation vector is tilted by an angle $\theta$ with respect to the cavity axis. 
    \textbf{b},~Schematic heterodyne spectrum of a levitated nanoparticle without cavity (grey) and filtered by a cavity (black).
    Six Lorentzian sidebands around the central carrier contain the information of the nanoparticle's three COM modes. The Stokes (anti-Stokes) amplitudes are proportional to $\bar{n}+1$ ($\bar{n}$), which we use for sideband thermometry. The cavity transfer function (dashed blue line) is detuned by $\Delta$ from the tweezers' frequency and enhances the anti-Stokes processes leading to cooling of the mechanical modes.}
    \label{fig:setup}
\end{figure}

A sketch of our optomechanical system is shown in figure~\ref{fig:setup}a, additional information can be found in the Methods. 
We detect and cool the COM mechanical modes of a single spherical SiO$_2$ nanoparticle of nominal diameter \SI[separate-uncertainty=true]{143(6)}{\nano\meter} and mass $\SI[separate-uncertainty=true]{3.4(4)}{\femto\gram}$. 
The nanoparticle is levitated in high vacuum (pressure of \SI[separate-uncertainty=true]{5(4)e-9}{\milli\bar}) using optical tweezers at a wavelength of $\SI[separate-uncertainty=true]{1550.0(5)}{\nano\meter}$ (frequency $\omega_0$) with optical power $\SI[separate-uncertainty=true]{1.20(8)}{\watt}$, focused by a high numerical aperture (NA $ = 0.75$) lens. The polarisation at the focus is defined by the tilt angle $\theta$ between the major axis of the polarisation ellipse and the cavity axis and by the degree of ellipticity. We choose the polarisation by tuning a set of waveplates, compensating for the birefringence of our vacuum window and trapping lens. The nanoparticle's reference frame is defined by the tweezers' propagation ($z$) and polarization ($x$) axes, as well as the axis orthogonal to the two ($y$). 
Strong focusing of the linearly polarised optical tweezers results in non degenerate, bare mechanical frequencies of the COM motion $\Omega_{x,y,z}/2\pi = (\num[separate-uncertainty = true]{224(2)}), (\num[separate-uncertainty = true]{268(2)}), \SI[separate-uncertainty = true]{80(1)}{\kilo\hertz}$.
The asymmetric cavity consists of two mirrors with different transmission separated by $\SI[separate-uncertainty = true]{6.4(1)}{\milli\meter}$, resulting in a linewidth of $\kappa/2\pi = \SI[separate-uncertainty = true]{330(9)}{\kilo\hertz}$. 
The nanoparticle scatters light into the cavity, which leaks through the higher transmission mirror, is combined with a local oscillator ($\omega_\mathrm{LO}/2\pi  =\omega_0/2\pi + \SI{1.5}{\mega \hertz}$) and then split equally onto a balanced photodetector. 
Heterodyne spectra are calculated as power spectral densities (PSDs) of the balanced photodetector voltage. 
\\

\centerline{\textbf{Cooling to 2D ground-state}}

\begin{figure*}
    \centering
    \includegraphics[]{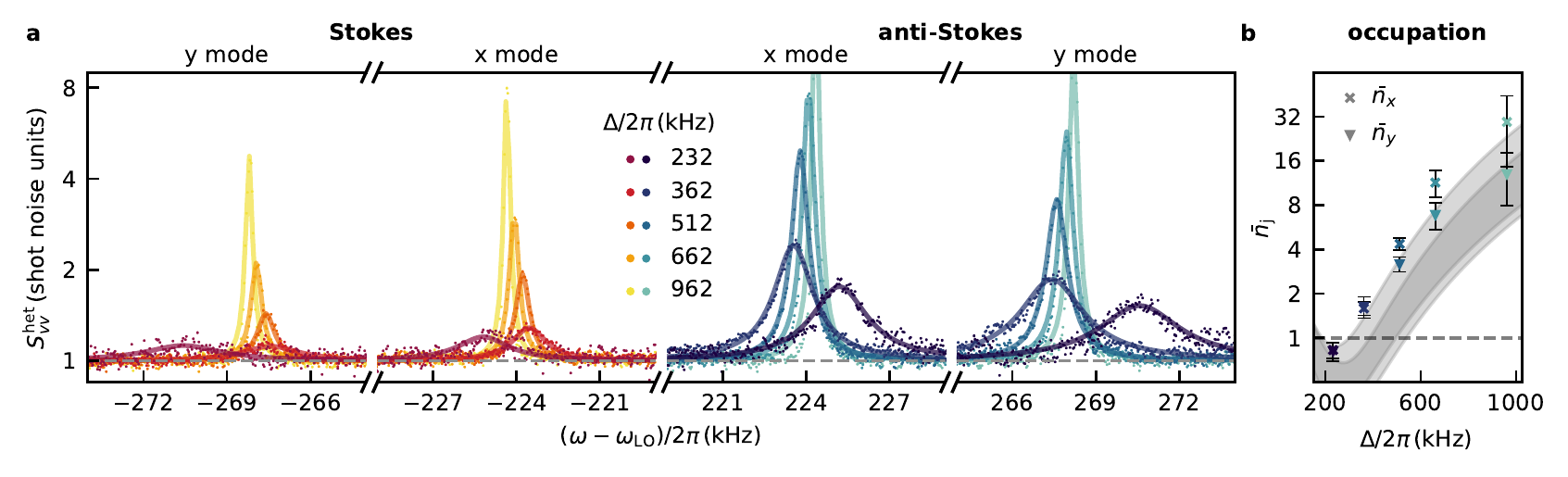}
    \caption{\textbf{Two-mode ground-state cooling.} \textbf{a},~Heterodyne PSDs of Stokes and anti-Stokes sidebands of $x$ and $y$ modes for different cavity detunings $\Delta$. 
    From Lorentzian fits (lines) the thermal occupation numbers are extracted via sideband thermometry. 
    \textbf{b},~Occupation numbers (calculated from fits in \textbf{a}) for $x$ and $y$ modes as function of cavity detuning. 
    For $\Delta/2\pi = 232\,\rm{kHz}$, close to the optimal value $(\Omega_x+\Omega_y)/2$ for simultaneous cooling, both occupation numbers are cooled below 1 (grey dashed line). 
    Error bars are standard deviations of fitted parameters. 
    Shaded areas correspond to theoretical estimations of $n_x$ (upper) and $n_y$ (lower, overlap is darker) based on coupling and heating rates and their uncertainties extracted from the measured PSDs.
    }
    \label{fig:2DGS}
\end{figure*}
Our two-mode ground state cooling experiment relies on coherent scattering~\cite{Hechenblaikner1998,Vuletic2000,Hosseini2017}, referring to light being scattered off a polarisable particle populating an optical cavity.
This method has gained significant interest as, compared to other cavity cooling schemes, it offers larger optomechanical coupling strengths and reduced phase noise heating~\cite{Delic2019}. 
Here, we exploit coherent scattering for coupling two motional modes of a single nanoparticle to an optical cavity mode~\cite{Windey2019,Gonzalez-Ballestero2019,Delic2019,Ranfagni2021}. 
A harmonically trapped nanoparticle scatters light elastically (Rayleigh) and inelastically (Raman). 
The Raman processes lead to sidebands in the scattered light spectrum. 
Figure~\ref{fig:setup}b shows a schematic of the resulting heterodyne spectrum which illustrates the cooling mechanism of coherent scattering.
The positive and negative frequencies correspond to the destruction and creation of a phonon, which are denoted by anti-Stokes and Stokes scattering, respectively. 
The grey line represents the spectrum of the mechanical oscillations without cavity. 
We introduce an optical cavity (dashed blue line is the intensity transfer function), whose resonance frequency $\omega_\mathrm{c}$ is detuned by $\Delta = \omega_\mathrm{c} - \omega_0$.
As we choose $\Delta \approx (\Omega_x+\Omega_y)/2$ the cavity enhances anti-Stokes relative to Stokes scattering in the spectrum (black line), which reduces the occupation numbers $\bar{n}_{j}$ ($j=x,y$) of the COM modes.
The asymmetry between Stokes and anti-Stokes peaks is additionally influenced by the fact that their scattering rates are proportional to $\bar{n}_j+1$ and $\bar{n}_j$, respectively. 
Taking into account the cavity transfer function, we use the measured asymmetry in the PSDs to extract $\bar{n}_{x,y}$ through a technique called sideband thermometry~\cite{Leibfried2003,Delic2019} (see Methods). 
Figure \ref{fig:2DGS}a shows the measured heterodyne PSDs normalised to shot noise level.
The PSDs contain Stokes and anti-Stokes sidebands of both transversal modes ($x$ and $y$), for different cavity detunings $\Delta$ at $\theta = 0.25 \pi$. 
The cooling by coherent scattering becomes more efficient as $\Delta$ approaches $(\Omega_x+\Omega_y)/2$ which results in a smaller amplitude and broader width of the sidebands. 
For each detuning and each COM mode, we fit Lorentzians (lines) of equal widths but independent amplitudes to the Stokes and anti-Stokes sidebands. 
The asymmetries which we use for sideband thermometry are then given by the ratio of anti-Stokes to Stokes amplitudes. 
Figure~\ref{fig:2DGS}b shows the extracted occupation numbers as a function of cavity detuning. 
The shaded areas represent simulations based on coupling strengths $g_j$ and heating rates $\Gamma_j$ which we extract (see Methods) from our data to be $g_x/2\pi  = \SI[separate-uncertainty = true]{14.1(27)}{\kilo\hertz}$, $g_y/2\pi  = \SI[separate-uncertainty = true]{15.4(19)}{\kilo\hertz}$, $\Gamma_x/2\pi  = \SI[separate-uncertainty = true]{1.0(4)}{\kilo\hertz}$ and $\Gamma_y/2\pi  = \SI[separate-uncertainty = true]{1.0(4)}{\kilo\hertz}$ . 
We find the heating rates to be limited by photon recoil (see Methods), as they are in good agreement with values calculated  from system parameters . 
For $\Delta/2\pi = \SI{232}{\kilo\hertz}$ close to $(\Omega_x+\Omega_y)/2$ we reach occupation numbers of $\bar{n}_{x}  = \num[separate-uncertainty = true]{0.83(10)}$ and $ \bar{n}_{y}  = \num[separate-uncertainty = true]{0.81(12)}$, %cooling both modes simultaneously to their quantum ground states. \vspace{5mm}
cooling the COM motion into its two-dimensional (2D) quantum ground-state. 
\\

\centerline{\textbf{Transition from 2D to 1D ground-state cooling}}

We explore the robustness of our cooling scheme to changes of the coupling rates by changing the polarisation angle $\theta$ of the trapping light.
The linearised optomechanical coupling strengths $g_{x,y}$ for linear polarisation have the form $g_x\propto\cos{\theta}$ and $g_y\propto\sin{\theta}$~\cite{Gonzalez-Ballestero2019}.
\begin{figure}
    \centering
    \includegraphics[]{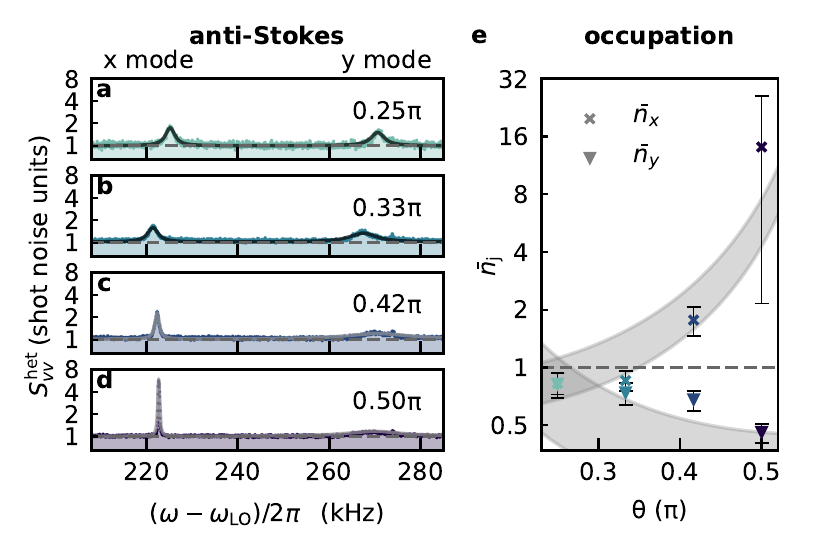}
    \caption{\textbf{Polarisation dependence of two-mode cooling.} \textbf{a-d},~Anti-Stokes sidebands of $x$ and $y$ modes and Lorentzian fits (lines) for different polarisation angles $\theta$. \textbf{a}, Both modes have similar coupling strengths for $\theta=0.25\pi$. \textbf{d}, Optimal $y$ axis cooling is achieved for $\theta = 0.5\pi$. \textbf{e},~Occupation numbers of $x$ and $y$ modes separate for $\theta > 0.25\pi$, transitioning from 2D to 1D cooling. Error bars reflect the standard deviation of the fitted parameters. Shaded areas mark theoretical predictions based on extracted coupling strengths.
    }
    \label{fig:lin_polarisation}
\end{figure}
Fig.~\ref{fig:lin_polarisation}a-d display anti-Stokes sidebands of $x$ and $y$ modes for different $\theta$ at $\Delta/2\pi = \SI[separate-uncertainty = true]{246(8)}{\kilo\hertz}$.
By tuning $\theta$ from $0.25\pi$ to $0.5\pi$, we observe the transition from 2D to 1D ground-state cooling. Close to $\theta=0.5\pi$ the shrinking/rising peak amplitudes indicate the motion along $y$/$x$ being cooled more/less efficiently due to larger/smaller coupling strength $g_y$/$g_x$.
Note that at $\theta=0.5\pi$ the $x$ motion is still imprinted in the spectrum of the cavity field and significantly cooled.
We attribute this to imperfections in the polarization state and in the angle alignment between the optical axes of trap and cavity. 
Additionally, small shifts in the frequencies $\Omega_j$ for different $\theta$ are caused by power drifts of the optical tweezers on the 5\% level. 
The extracted phonon occupations from sideband thermometry $\bar{n}_{x,y}$ are displayed in Fig.~\ref{fig:lin_polarisation}e.
The simulations (shaded areas, see Methods) show the increasing and decreasing occupation numbers of the $x$ and $y$ mode, respectively, in agreement with the data.
This result is well aligned with the predicted decrease (increase) of $g_x\propto\cos{\theta}$($g_y\propto\sin{\theta}$) as $\theta$ is increased from $0.25\pi$ to $0.5\pi$.
Experimentally, we find robust two-mode ground-state cooling at $\theta=0.25\pi$ and $0.33\pi$. Furthermore, we observe our lowest single-mode phonon occupation $\bar{n}_{y}  = \num[separate-uncertainty = true]{0.46(5)}$ paired with a high phonon occupation $ \bar{n}_{x}  = \num[separate-uncertainty = true]{14(12)}$ at $\theta=0.5\pi$.
\\

\centerline{\textbf{Limits of 2D sideband thermometry}}

To efficiently cool two COM modes ($x, y$) of a levitated nanoparticle, several conditions must be met. First, the optical cavity must simultaneously resolve the anti-Stokes sidebands of the $x$ and $y$ modes, i.e. $\vert \Omega_y- \Omega_x\vert \lesssim \kappa \lesssim  \Omega_y, \Omega_x$.
Further, the system needs to be in the weak coupling regime $\vert g_j \vert \ll \kappa$, in order to prevent hybridisation of the cavity and mechanical modes, which hinders efficient cooling~\cite{aspelmeyer2014cavity}. 
Finally, $\Omega_x$ and $\Omega_y$ must be sufficiently separated. $x$ and $y$ modes are cooled by the cavity via a collective bright mechanical mode, while the orthogonal dark mechanical mode is only sympathetically cooled when coupling to the bright mode~\cite{Genes2008}. For near-degenerate $\Omega_x$ and $\Omega_y$, the dark mode decouples and this inhibits cooling of its constituent $x$ and $y$ modes (see Methods).
The condition $\vert \Omega_y- \Omega_x\vert \gtrsim \vert g_j \vert$ is thus necessary for 2D ground-state cooling~\cite{Genes2008, Toros2021}.

\begin{figure*}
    \centering
    \includegraphics[width = \textwidth]{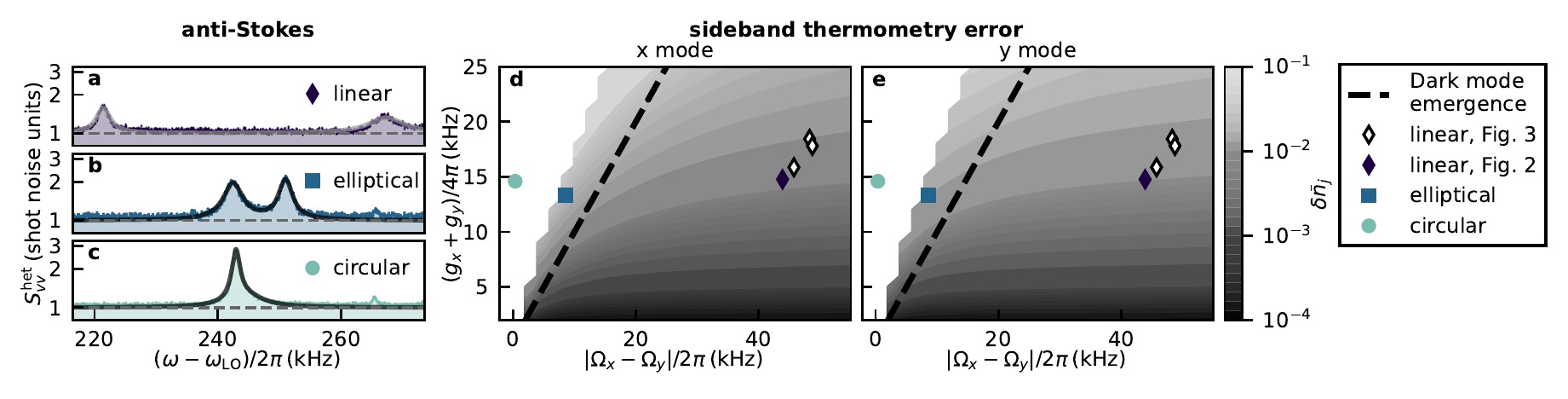}
    \caption{\textbf{Limits of 2D sideband cooling.} \textbf{a-c},~Anti-Stokes sidebands of $x$ and $y$ modes for different degrees of ellipticity in the trap's polarisation.
    The separation between peaks $\Omega_y-\Omega_x$ decreases with increasing ellipticity. 
    In \textbf{b}, cross-coupling increases the occupation numbers to $\bar{n}_{x}  = \SI[separate-uncertainty = true]{2.1(4)}{}$ and $ \bar{n}_{y}  = \SI[separate-uncertainty = true]{2.8(8)}{}$. 
    Circular polarisation causes degenerate peaks in \textbf{c}.
    \textbf{d-e},~Relative error of $x$ (\textbf{d}) and $y$ (\textbf{e}) occupation extracted by sideband thermometry to real phonon number. 
    Parameter pairs of measurements in figures~\ref{fig:2DGS} ,~\ref{fig:lin_polarisation} and~\ref{fig:limits} are marked by points and their corresponding error of few percent added to all presented phonon numbers. 
    The dark mode decoupling prevents efficient cooling for $\Omega_y-\Omega_x<g$ (black line) and sideband thermometry becomes impossible for degenerate peaks (white area).
    }
    \label{fig:limits}
\end{figure*}
The unique in-situ tunability of levitated systems allows us to observe the effect of the dark mode decoupling on $\bar{n}_{x,y}$.
We change $\Omega_{x,y}$ by tuning the ellipticity of the trapping beam while keeping $\theta = 0.25\pi$ and $\Delta/2\pi = \SI[separate-uncertainty = true]{257(11)}{\kilo\hertz}$. 
Comparing Fig.~\ref{fig:limits}a and b we observe that $\Omega_{x}$ and $\Omega_{y}$ approach each other as the polarization changes from linear to elliptical. 
In Fig.~\ref{fig:limits}b both modes heat up to $\bar{n}_{x}  = \num[separate-uncertainty = true]{2.0(4)}$ and $ \bar{n}_{y}  = \num[separate-uncertainty = true]{2.8(8)}$ as weak coupling of the dark mode inhibits cooling. 
As we polarise the tweezers circularly for Fig~\ref{fig:limits}c, $x$ and $y$ peaks merge and we are unable to extract individual occupation numbers by conventional sideband thermometry.

We further theoretically test the validity of extracting phonon numbers by sideband thermometry using a full quantum model~\cite{Gonzalez-Ballestero2019} (see Methods). 
We first calculate the true phonon occupation $\bar{n}^{\text{model}}_{j}$ for $j=x,y$. 
Then, using the same model, we calculate PSDs of the heterodyne detection and perform sideband thermometry on them to extract $\bar{n}_j$.
We define the systematic error $\delta \bar{n}_j = \vert(\bar{n}^{\text{model}}_j -\bar{n}_j)/\bar{n}^{\text{model}}_j\vert$ and show the result in Fig.~\ref{fig:limits}d,e.
Mostly we find that $\bar{n}_j$ underestimates $\bar{n}^{\text{model}}_j$.
The error is minimal for well separated COM mode frequencies in the weak coupling regime. 
For stronger coupling and constant mode spacing, hybridisation between optical and mechanical modes becomes more significant and $\delta \bar{n}_j$ increases.
At constant coupling rate, the error also increases as the mechanical frequencies approach degeneracy and the effect of the dark mode gains importance.
We can not perform 2D sideband thermometry for degenerate peaks, which occurs at small mode spacing and large coupling strength (white areas).
Finally we display the estimated errors for all measurements presented in Figs.~\ref{fig:2DGS}-\ref{fig:limits}.
These errors of our sideband thermometry method are marginal, for Fig.~\ref{fig:2DGS} only about 1\%, which certifies that we have achieved two-mode ground-state cooling.
\\

\centerline{\textbf{Conclusions}}

We have simultaneously prepared two out of three centre-of-mass modes of a levitated particle in their ground state with residual occupation numbers of $\bar{n}_{x}=0.83$ and $\bar{n}_{y}=0.81$. 
With respect to the optical axis of the tweezer, our cooling scheme controls the transversal degrees of freedom, resulting in two important implications.

First, together with the demonstrated ground-state cooling along the tweezers' axis~\cite{Magrini2021, Tebbenjohanns2021,Kamba2022}, 3D centre-of-mass quantum control is within experimental reach.  
Demonstrating three dimensional ground-state cooling would be a significant step towards full control of large systems at the quantum limit.

Second, control over transversal centre-of-mass motion implies control of the orbital angular momentum along the tweezers' axis, given by $\hat{L}_z = \hat{x}\hat{p}_y - \hat{y}\hat{p}_x$, where $(\hat x, \hat y)$ and $(\hat p_x,\hat p_y)$ are the transverse position and momentum vector operator, respectively.
As our transversal motion is in a thermal state, the variance of the corresponding angular momentum is given by $\langle \hat{L}_z^2 \rangle/\hbar^2= (\bar{n}_{x}+1/2) (\bar{n}_{y}+1/2) (\Omega_x/\Omega_y + \Omega_y/\Omega_x)-1/2$. 
With our occupation numbers and trap frequencies, we find $\sqrt{\langle \hat{L}_z^2 \rangle}\approx1.7\,\hbar$.
We have therefore prepared our system close to an angular momentum eigenstate along $z$ ($\langle \hat{L}_z^2 \rangle =0$) with $\langle \hat{L}_z \rangle =0$. 
This opens the door to realising protocols combining 2D ground-state cooling with coherently pumped orbital angular momentum~\cite{Svak2018} to stabilise a state with large orbital angular momentum $\langle \hat{L}_z \rangle \gg \hbar$ and quantum-limited variance. 
Those minimally fluctuating high orbital angular momentum states (``quantum orbits'') would not only be promising for fundamental studies of low-noise and massive high angular momentum states, but also for becoming building blocks of a gyroscope with quantum-limited performance.

\emph{Data availability} The datasets generated and analysed during the current study will be made available in the ETH Zurich Research Collection prior to publication.\\
\emph{Author contributions} J.~P., D.~W. and J.~V. conducted the experiments, C.~G.~B. and O.~R.~I. performed the theoretical modelling, A.~d.~l.~R.~S., N.~M. and R.~Q. conceptualised the setup with R.~R. and L.~N., who directed the project.\\
\emph{Acknowledgements} This research has been supported by the European Research Council (ERC) under the grant Agreement No. [951234] (Q-Xtreme ERC-2020-SyG) and by the European Union’s Horizon 2020 research and innovation programme under grant no. 863132 (iQLev).
We thank our colleagues E.~Bonvin, L.~Devaud, M.~Frimmer, J.~Gao, M.~L.~Mattana, A.~Militaru, M.~Rossi, N.~Carlon Zambon and J.~Zielinska for input and discussions.

\FloatBarrier
\bibliographystyle{apsrev4-1}
\bibliography{Main}

\end{document}

\scalefont{1.0}

\title{Methods}

\author{Johannes~\surname{Piotrowski} }
\email[equal contribution]{ }
\affiliation{Photonics Laboratory, ETH Z{\"u}rich, 8093 Z\"urich, Switzerland}
\author{Dominik~\surname{Windey}}
\email[equal contribution]{ }
\affiliation{Photonics Laboratory, ETH Z{\"u}rich, 8093 Z\"urich, Switzerland}
\author{Jayadev~\surname{Vijayan}}
\affiliation{Photonics Laboratory, ETH Z{\"u}rich, 8093 Z\"urich, Switzerland}
\author{Carlos~\surname{Gonzalez-Ballestero}}
\affiliation{Institute for Quantum Optics and Quantum Information, Austrian Academy of Sciences, A-6020 Innsbruck, Austria}
\affiliation{Institute for Theoretical Physics, University of Innsbruck, A-6020 Innsbruck, Austria}
\author{Andr\'es  ~\surname{de los R\'ios Sommer}}
\affiliation{Nanophotonic Systems Laboratory, ETH Z\"urich, 8092 Z\"urich, Switzerland}
\author{Nadine~\surname{Meyer}}
\affiliation{Nanophotonic Systems Laboratory, ETH Z\"urich, 8092 Z\"urich, Switzerland}
\author{Romain~\surname{Quidant}}
\affiliation{Nanophotonic Systems Laboratory, ETH Z\"urich, 8092 Z\"urich, Switzerland}
\affiliation{Quantum Center, ETH Z\"urich, 8083 Z\"urich, Switzerland}
\author{Oriol ~\surname{Romero-Isart}}
\affiliation{Institute for Quantum Optics and Quantum Information, Austrian Academy of Sciences, A-6020 Innsbruck, Austria}
\affiliation{Institute for Theoretical Physics, University of Innsbruck, A-6020 Innsbruck, Austria}
\author{Ren\'e \surname{Reimann}}
\affiliation{Quantum Research Centre, Technology Innovation Institute, Abu Dhabi, UAE}
\author{Lukas~\surname{Novotny}}
\affiliation{Photonics Laboratory, ETH Z{\"u}rich, 8093 Z\"urich, Switzerland}

\date{\today} % Leave empty to omit a date
\maketitle
\centerline{\textbf{Setup}}

\begin{figure*}
    \centering
    \includegraphics[width = 0.9\linewidth]{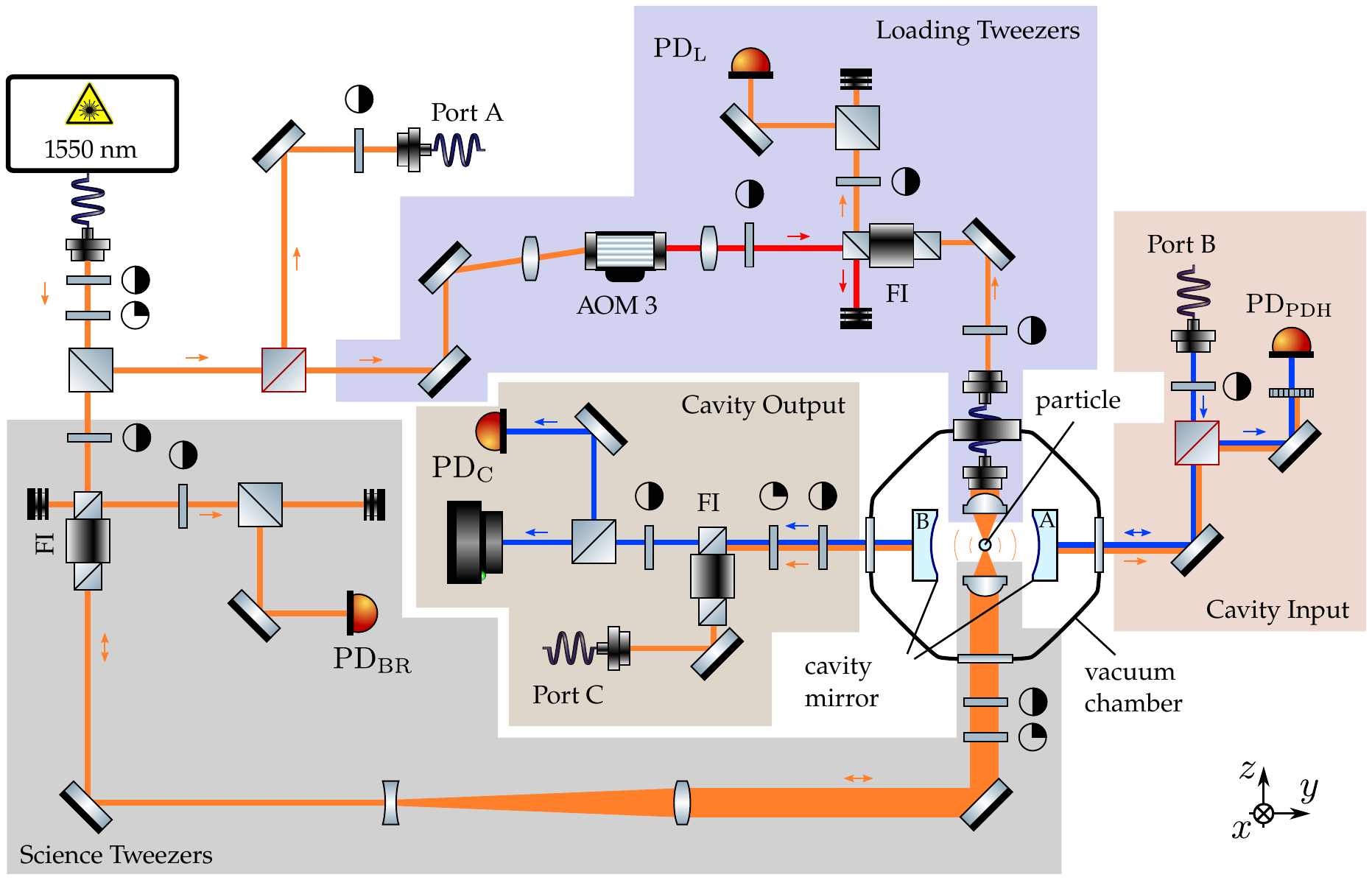}
    \caption{\textbf{Core setup for particle trapping, transfer and detection} To simplify the sketch, we show components of the detection setup and cavity lock on a separate sketch in Fig.~\ref{fig:setup_B}. 
    We link the ports as indicated by the letter. All beam splitters with a black (red) outline are polarizing (non-polarizing) beam splitters and components labeled FI are Faraday isolators. All beams are derived from a \emph{NKT Photonics E15} $\SI{1550}{\nano\meter}$ laser. In the sketched configuration (particle loaded in science tweezers), the half wave plate in the loading tweezers' section is set to dump all power at the input of a FI. Initially, while loading the particle, the loading tweezers are positioned in a separate vacuum chamber (not shown) and the full power is used to trap a particle. The photodetector $\mathrm{PD}_\mathrm{L}$ is used to monitor the trapping process. After aligning the loading tweezers with the science tweezers we rotate the half wave plate in front of the FI in the science tweezers' section to turn on the science tweezers. At the same time, we rotate the half-wave plate in the loading tweezers section, to turn off the loading tweezers and transfer the particle. From port B we feed in a beam to lock the cavity length by using the signal of the photodetector $\mathrm{PD}_\mathrm{PDH}$. To reduce noise on the detector we cross polarize the beam w.r.t to the tweezers and use a polarizer to filter out the particle scattered light. On the opposite side of the cavity we use a photodiode to monitor the lock quality $\mathrm{PD}_\mathrm{c}$ and an infrared camera to image the cavity mode. The mirror on the right is the high finesse mirror, therefore most of the particle scattered light leaks through the left mirror. We feed the light that leaks out of the cavity to port C to detect it with the heterodyne setup.
    }
    \label{fig:full-setup}
\end{figure*}
\begin{figure}
    \centering
    \includegraphics[width = \columnwidth]{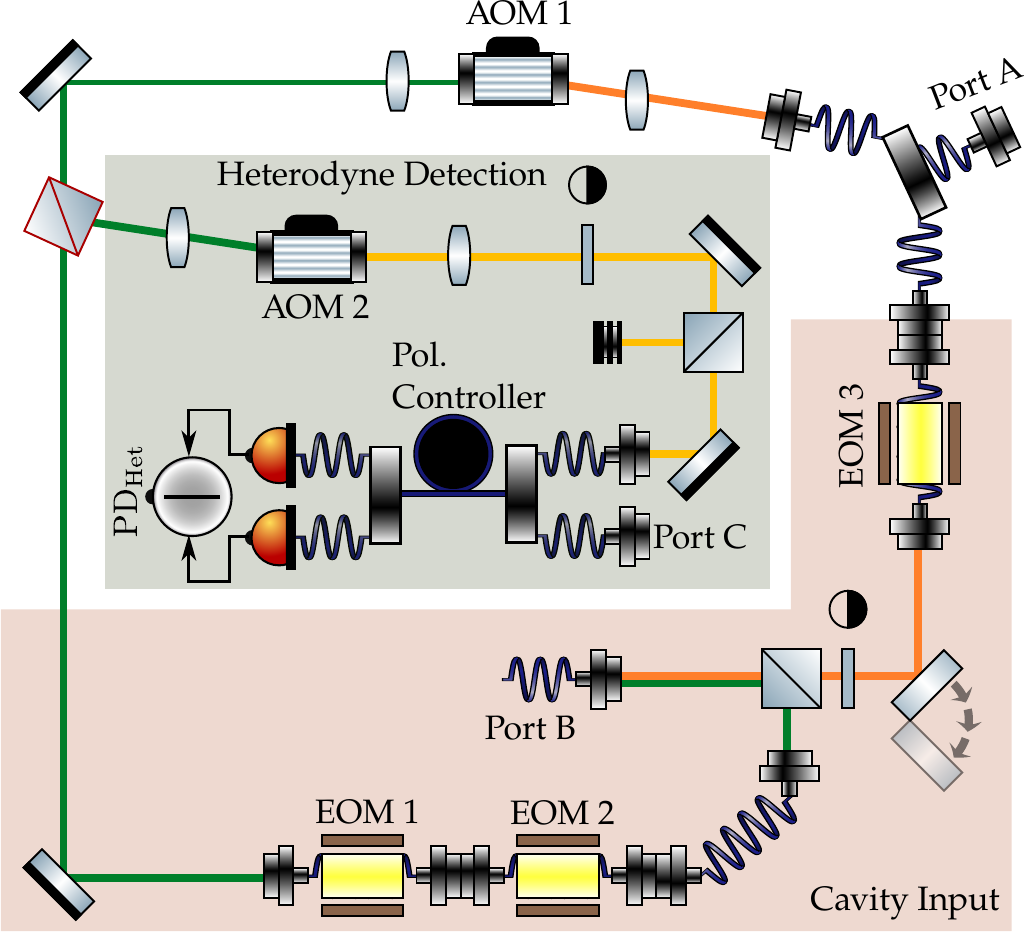}
    \caption{\textbf{Constituent setup for particle detection, cavity locking and calibration.} All components are sketched as described in Fig.~\ref{fig:full-setup}. Light from the core setup enters from the top right through port A. We drive AOM1 ($+1\mathrm{st}$ order) and AOM2 ($-1\mathrm{st}$ order) at $2\pi \times \SI{80}{\mega\hertz}$ and $2\pi \times \SI{78.5}{\mega\hertz}$, resulting in a local oscillator detuning $\Delta_\mathrm{lo} = \SI{1.5}{\mega\hertz}$. As we lock to the $\mathrm{TEM}_{10}$ mode of the cavity, we use AOM1 and EOM1 to derive a beam at frequency close to $\omega_{10} - \omega_{00}$ (the difference of the resonance frequencies of $\mathrm{TEM}_{10}$ and $\mathrm{TEM}_{00}$). On the opposite side of the cavity input section, we modulate sidebands on the calibration beam before combining it with the lock beam. We implement a flip mirror to prevent the calibration beam from entering port B and consequently the cavity, while doing measurements.
    }
    \label{fig:setup_B}
\end{figure}
A detailed view of the experimental setup is shown in Fig.~\ref{fig:full-setup} and Fig.~\ref{fig:setup_B}.
To keep our cavity free of contaminants and at vacuum conditions (below $\SI{e-2}{\milli\bar}$) at all times, we load nanoparticles (microParticles GmbH, SiO2-F-L3205-23, \SI[separate-uncertainty=true]{143(6)}{\nano\meter} nominal diameter) in a separate loading chamber (not drawn in Fig.~\ref{fig:full-setup}) using a nebuliser (Omron). The loading tweezers are mounted on a movable arm and its light is frequency shifted by an acousto-optic modulator (AOM3) by \SI{80}{\mega\hertz} to avoid interference. After loading, we seal and evacuate the loading chamber to \SI{1e-2}{\milli\bar} and move the loading tweezers into the cavity chamber. We use two photodetectors and the loading tweezers' nanopositioner (SmarAct GmbH) to align the focal points of loading and science tweezers. Initially, we measure the intensity of light passing by the particle on $\mathrm{PD}_\mathrm{BR}$ to roughly align the two foci, while the science tweezers are still turned off. Afterwards, we gradually increase the power of the science tweezers and measure the light coupling into the fibre of the loading tweezers and shining on $\mathrm{PD}_\mathrm{L}$. 
Eventually, we turn down and up the power in the trapping and science tweezers, respectively, to transfer the nanoparticle to the science tweezers. Thereafter, the science chamber is sealed off and pumped down to \SI{5e-9}{\milli\bar} with a combination of turbomolecular pump ion-getter pump. 

The cavity with linewidth $\kappa/2\pi = \SI{330}{\kilo\hertz}$, Finesse $\mathcal{F} = 70\mathrm{\,}\mathrm{k}$ and focal spot waist $w_\mathrm{c} = \SI[separate-uncertainty=true]{48(5)}{\micro\meter}$ is built from a low and high Finesse mirror (specified Finesse of coatings: 37k and 126k) with a radius of curvature $\mathrm{ROC} = \SI{10}{\milli\meter}$ and cavity length $L = \SI{6.4}{\milli\meter}$. The absorption $A$ and transmission $T$ of the low (high) Finesse coatings are $A, T = 4, 79 \mathrm{\,ppm}$ ($A, T = 5, 20 \mathrm{\,ppm} $). We lock the cavity to the $\mathrm{TEM}_{10}$ mode of the cavity to avoid interference with light scattered off the particle into the $\mathrm{TEM}_{00}$ mode. The necessary frequency shift $\omega_{10} - \omega_{00}$ of approximately $\SI{8}{\giga\hertz}$ is generated by AOM1 and EOM1 to obtain the lock beam sent into the cavity. We define the cavity detuning $\Delta = \omega_\mathrm{c} - \omega_0$ with respect to the cavity resonance frequency $\omega_\mathrm{c}$ of the $\mathrm{TEM}_{00}$ mode. The backscattered light is collected on a photodiode ($\mathrm{PD}_\mathrm{PDH}$) to generate the error signal for a Pound-Drever-Hall lock~\cite{Drever1983} of the cavity length. The small sidebands necessary for the error signal are generated by EOM2 at $\SI{23}{\mega\hertz}$. 
\\

\centerline{\textbf{Detuning calibration}}

\begin{figure}
    \centering
    \includegraphics[]{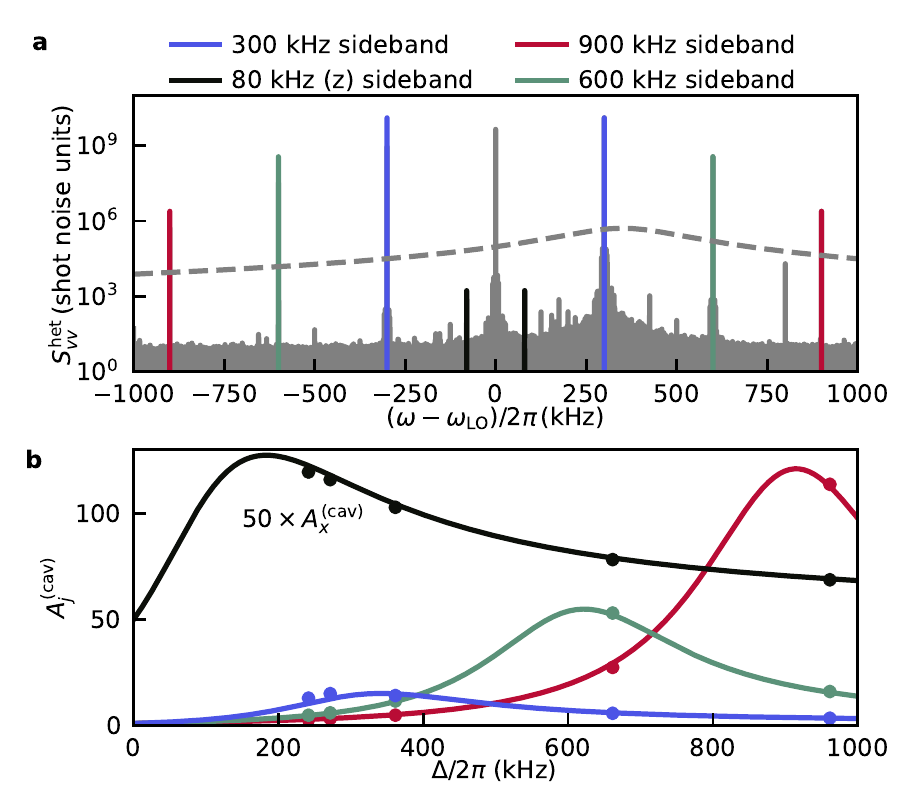}
    \caption{\textbf{Detuning calibration by known sidebands.} \textbf{a,} EOM-modulated sidebands of the calibration laser at \SI{900}{\kilo\hertz} (red), \SI{600}{\kilo\hertz} (green) and \SI{300}{\kilo\hertz} (blue), as well as the uncooled $z$-peak at \SI{80}{\kilo\hertz} (black) are filtered by the cavity transfer function (grey dashed line). \textbf{b,} We fit (lines) the detuning-dependent, expected cavity-induced asymmetry $A_{j}^\text{(cav)}$ to the measured asymmetries at of all four sidebands. The asymmetry of the $x$-peak is multiplied by $50$ for visibility.
    }
    \label{fig:calib}
\end{figure}
To perform accurate sideband thermometry we need to compensate for the effect of cavity filtering of the motional sidebands. For a known detuning $\Delta$ and linewidth $\kappa$ we can readily calculate the cavity-induced asymmetry at a mode frequency $\Omega_j$ by assuming a Lorentzian filter function~\cite{Delic2020}:
\begin{equation*}
  A_{j}^\text{(cav)}=\frac{\kappa^{2}+4\left(\Delta+\Omega_{j}\right)^{2}}{\kappa^{2}+4\left(\Delta-\Omega_{j}\right)^{2}} \;.
\end{equation*}
In our experiments the particle scatters light into the $\mathrm{TEM}_{00}$ of the optical cavity, while the cavity is locked to the $\mathrm{TEM}_{10}$ mode. The difference of the two $\mathrm{TEM}$ mode frequencies is very sensitive to drifts of the cavity length between experiments. Thus we need to calibrate the detuning of the cavity with respect to the science tweezers.
As a particle-independent measure we temporarily send a calibration laser through the cavity before performing cooling experiments. This laser has sidebands modulated by an electro-optic modulator (EOM3) at \SI{300}{\kilo\hertz} and higher harmonics, which are filtered by the detuned cavity and then detected in our heterodyne scheme, as shown in Fig.~\ref{fig:calib}a. We extract the asymmetry of these three sidebands and the $z$-peak of the particle motion while tuning the cavity lock frequency with EOM1. In Fig.~\ref{fig:calib}b we fit the Lorentzian cavity filter function to the measured sideband asymmetry to extract the linewidth and calibrate the detuning $\Delta$. The linewidth agrees with an independent measurement of $\kappa/2\pi = \SI{330}{\kilo\hertz}$. The standard deviation for $\Delta$ in Fig. 2a are below \SI{3}{\kilo\hertz} and therefore the error bars are smaller than the markers.
\\

\centerline{\textbf{Sideband Thermometry}}

For extracting the phonon occupation from measured heterodyne signals we rely on the different scaling of Stokes and anti-Stokes scattering processes. As the latter requires
the presence of a phonon, its scattering rate
scales with the average phonon number $n$ while the Stokes scattering rate
scales with $n + 1$. This leads to an asymmetry of the anti-Stokes and Stokes sidebands dependent on the occupation number:
\begin{equation*}
    A_j^{(n)} = \frac{\bar{n}_j}{\bar{n}_j+1}\;.
\end{equation*}
The total asymmetry of the PSD at the frequency $\Omega_j$:
\begin{equation*}
    A_j = \frac{S_{vv}(\Delta_\text{lo}+\Omega_j)}{S_{vv}(\Delta_\text{lo}-\Omega_j)} = A_j^{(n)}A_j^\text{(cav)}
\end{equation*}
is the product of the thermal asymmetry $A_j^{(n)}$ and the cavity induced asymmetry $A_j^\text{(cav)}$. 
Given both, we can calculate the average occupation number
\begin{equation*}
\bar{n}_j=\frac{A_j}{A_j^{(\text{cav})}-A_j}\;.
\end{equation*}
To access the thermal asymmetry we assume a Lorentzian shape of the motional sidebands 
\begin{equation*}
S_{jj}(\Omega) \approx \frac{a_j}{2 \pi} \frac{\frac{\gamma_j}{2}}{\left(\Omega-\Omega_j\right)^2+\left(\frac{\gamma_j}{2}\right)^2}\;,
\end{equation*}
with amplitude $a_j$, width $\gamma_j$ and centre frequency $\Omega_j$. 
Our fitting function
\begin{equation*}\begin{aligned}
    F =  S_{xx,(\text{AS})}+S_{xx,(\text{S})}+  S_{yy,(\text{AS})}+S_{yy,(\text{S})} +S_{\text{SN}}
\end{aligned}\end{equation*}
lets us extract four values for $a_j$, $\gamma_j$ and $\Omega_j$, one for each Stokes and anti-Stokes sideband of mode $j=x,y$ and gives $A_j^{(n)} = a_{j,\text{AS}}/a_{j,S}$. We require $\gamma_{j,\text{AS}}= \gamma_{j,\text{S}}$ and $\Omega_{j,\text{AS}}= -\Omega_{j,\text{S}}$. The shot noise level $S_{\text{SN}}$ is extracted for each PSD in a region far away from any spectral features to account for small drifts in the local oscillator power. We normalise all spectra to the shot noise level. 
For the error bars we propagate the standard deviations of the fitted amplitudes $a_j$ and the cavity parameters $\Delta$ and $\kappa$.
\\

\centerline{\textbf{Quantum model}}

The Hamiltonian describing CS is \cite{Gonzalez-Ballestero2019} 
$$ \frac{\hat{\rm H}_{\text{CS}}}{\hbar}= \Delta\hat{a}^\dagger\hat{a} + \sum_{j =x,y,z} \Omega_j\hat{b}^\dagger_j\hat{b}_j- \sum_{j =x,y,z} (g_{j}\hat{a}^\dagger + \text{H.c.})  (\hat{b}_j^\dagger + \hat{b_j}),  $$
with $\hat{a}$ ($\hat{a}^\dagger$) being the photon annihilation (creation) operator and  $\hat{b}_j$ ($\hat{b}_j^\dagger$) the phonon annihilation (creation) operator along motional axis $j=x,y,z$. 
This interaction allows to cool the COM motion along all three axes, as has been demonstrated experimentally \cite{Windey2019,Delic2019}. 
The linearised optomechanical coupling strengths $g_{j}$ for linear polarisation are given by
\begin{equation}
	\begin{bmatrix} \label{eq:gj}
		g_x\\
		g_y\\
		g_z
	\end{bmatrix}
	=  - \frac{G_0}{2}\begin{bmatrix}
		 \:k_\text{c} \: x_{\text{zpf}} \:\sin{\phi}  \cos{\theta} \\
		 \:k_\text{c} \: y_{\text{zpf}} \:\sin{\phi} \sin{\theta} \\
		-i \: k_\text{t} \: z_{\text{zpf}} \:\cos{\phi}\\
	\end{bmatrix} 
\end{equation}
with $k_\text{c} = 2\pi/\lambda_\text{c}$ the cavity wavevector,  $[x_{\text{zpf}},y_{\text{zpf}},z_{\text{zpf}}] =  \sqrt{\frac{\hbar}{2\: m\: \Omega_{x,y,z}}}$ the zero-point fluctuations along each axis, and $\phi = 2\pi y_0/\lambda_\text{c}$, with $y_0$ being the particle position along the cavity axis and $y_0 = \lambda_\text{c}/4$ corresponding to an intensity minimum of the cavity standing wave.
The rate $G_0 = \alpha E_0 \sqrt{\frac{\omega_\text{c}}{2\hbar \epsilon_0 V_\text{c}}} \vec{e}_x\cdot\vec{e}_\alpha$ contains the particle polarizability $\alpha = 4\pi \epsilon_0 R^3 \frac{n_\text{r}^2-1}{n_\text{r}^2 +2}$, with $n_\text{r}$ the refractive index of the particle, $R$ its radius, and $\epsilon_0$ the vacuum permittivity, the trap electric field amplitude $E_0 = \sqrt{\frac{4 P}{\pi \epsilon_0 c \text{w}_\text{x} \text{w}_\text{y}}}$ with $\text{w}_\text{x,y}$ the trap waists at the focus, and the cavity parameters, namely mode volume $V_\text{c} = \pi \text{w}_\text{c}^2 L_\text{c}/4$, cavity waist $\text{w}_\text{c}$, cavity length $L_\text{c}$, and frequency $\omega_\text{c} = 2\pi c/\lambda_\text{c}$. 
Note that the dot product of vectors $\vec{e}_{x}\cdot\vec{e}_{\alpha}$ introduces an additional dependency on the trap polarization angle $\propto\cos\theta$, that hinders the possibility of coupling solely the $x$-motional mode to the cavity. 

The quantum state of the cavity-nanoparticle system is given by its density matrix $\hat{\rho}$, which obeys the dynamical equation~\cite{Gonzalez-Ballestero2019}
\begin{multline}\label{MasterEq}
    \frac{d}{dt}\hat{\rho} = -\frac{i}{\hbar}\left[\hat{H}_{\rm CS},\hat{\rho}\right] + \kappa \left[\hat{a}\hat{\rho}\hat{a}^\dagger - \{\hat{a}^\dagger\hat{a},\hat{\rho}\}/2\right]
    \\
    -\sum_{j=x,y,z}\frac{\Gamma_j}{2}\left[\hat{b}_j+\hat{b}^\dagger_j,\left[\hat{b}_j+\hat{b}^\dagger_j,\hat{\rho}\right]\right]
    \\+ \frac{\gamma}{4}\sum_{j=x,y,z}\left[\hat{b}_j+\hat{b}^\dagger_j,\left\{\hat{b}_j^\dagger-\hat{b}_j,\hat{\rho}\right\}\right]\;,
\end{multline}
where $\{*\}$ denotes the anticommutator, $\gamma$ is the friction rate due to gas damping, and the heating rates $\Gamma_j =\Gamma_j^{(r)}+\Gamma_j^{(g)}$ contain a contribution from gas molecules,
\begin{equation}\label{gammagas}
    \Gamma_j^{(g)} = \gamma\frac{k_B T}{\hbar\Omega_j}
\end{equation}
and a contribution from laser recoil heating,
\begin{equation}\label{gammarecoil}
    \left[
    \begin{array}{c}
         \Gamma_x^{(r)}  \\
         \Gamma_y^{(r)} \\
         \Gamma_z^{(r)}
    \end{array}
    \right] = \frac{\pi}{15\hbar\epsilon_0}\left(\frac{\alpha E_0}{2\pi}\right)^2 k_0^5 \left[
    \begin{array}{c}
         x_{\rm zpm}^2  \\
         2y_{\rm zpm}^2 \\
         7z_{\rm zpm}^2
    \end{array}
    \right].
\end{equation}
To obtain analytical expressions we simplify the model by making two assumptions. First, friction due to gas molecules is negligible, which at the pressures used for this work can be checked to be a good approximation~\cite{Gonzalez-Ballestero2019,Ranfagni2022}. This amounts to neglecting the friction term in the master equation, namely the last line in Eq.~\eqref{MasterEq}. Note that the associated heating rate $\Gamma_j^{(g)} \gg \gamma$ is not neglected. Second, the particle equilibrium position is at the intensity minimum of the cavity mode ($\phi = \pi/2$). This is the case for all measurements in this work within an accuracy of $\SI{1}{\nano \meter}$. At this position the couplings $g_{x,y}$ are simultaneously maximised while the $z$ mode becomes uncoupled. %, $g_z=0$. 
Under these approximations the system is reduced to a three-mode system including only the cavity mode and the $x$ and $y$ motional modes, whose steady-state properties can be computed analytically (see below). 

Within this three-mode approximation we can explain the dark-mode effect. The Hamiltonian can be cast in the form
\begin{multline}\label{Hbrightdark}
    \frac{\hat{\rm H}_{\text{CS}}}{\hbar}=
        \Delta\hat{a}^\dagger\hat{a} + \sum_{j =\pm}\omega_j\hat{B}_j^\dagger\hat{B}_j+ G_\pm \left(\hat{B}_+^\dagger\hat{B}_-+\text{H.c.}\right)
        \\-(g_t\hat{a}^\dagger+\text{H.c.})(\hat{B}_+^\dagger+\hat{B}_+)\;.
\end{multline}
Here, we have changed basis in the $x-y$ subspace to define two collective mechanical modes, namely the bright and dark modes
\begin{equation}
    \hat{B}_+ \equiv \frac{g_x\hat{b}_x + g_y\hat{b}_y}{g_t} \hspace{0.3cm};\hspace{0.3cm}  \hat{B}_- \equiv \frac{g_y\hat{b}_x - g_x\hat{b}_y}{g_t}\;,
\end{equation}
with the total coupling rate $g_t = \sqrt{g_x^2+g_y^2}$. The corresponding mechanical frequencies of bright and dark mode read
\begin{equation}
    \omega_+ = \frac{\Omega_xg_x^2+\Omega_yg_y^2}{g_t^2} \hspace{0.3cm};\hspace{0.3cm} \omega_- = \frac{\Omega_xg_y^2+\Omega_yg_x^2}{g_t^2}\;.
\end{equation}
According to the Hamiltonian Eq.~\eqref{Hbrightdark} the cavity couples directly only to the bright mode and can thus only cool this mode. The dark mode can be only sympathetically cooled through its coupling to the bright mode, which has a rate
\begin{equation}
    G_\pm = \frac{g_xg_y}{g_t^2}(\Omega_x-\Omega_y)\;.
\end{equation}
If the cavity, via the optomechanical coupling, cannot resolve the degeneracy between the two modes, i.e. if $g_{x,y} \gtrsim \vert \Omega_x-\Omega_y\vert$, the above coupling becomes much smaller than all other rates, $\Delta, \omega_{\pm},$ and $g_t$, preventing the dark mode from being effectively cooled. This in turn limits the steady-state occupations of the original modes which are now limited by the thermal dark-mode occupation, $\langle \hat{b}_{x,y}^\dagger\hat{b}_{x,y}\rangle \sim \langle \hat{b}_{-}^\dagger\hat{b}_{-}\rangle$.
\\

\centerline{\textbf{Extracting coupling and heating rates}}

Within the two above assumptions (negligible friction, $z-$motion uncoupled) the heterodyne power spectral density (PSD) can be analytically calculated.
The heterodyne PSD, after subtraction of the noise floor and normalization to it, can be written as~\cite{bowen2015quantum}
\begin{equation}\label{eq:heterodyne}
    S_{\rm het}(\omega) =  \kappa\left[S_{\rm c}(\delta_{\rm LO}-\omega)+S_{\rm c}(\omega+\delta_{\rm LO})\right]
\end{equation}
with $\delta_{\rm LO} = \omega_{\rm LO} - \omega_0$. It is expressed in terms of the cavity PSD,
\begin{equation}
    S_c(\omega) = \int_{-\infty}^\infty\frac{ds}{2\pi}e^{-i\omega s}\langle\hat{a}^\dagger(0)\hat{a}(s)\rangle_{\rm ss}\;,
\end{equation}
where the sub-index ``ss'' indicates the steady state. We compute the two-time correlator using the quantum regression theorem~\cite{Gonzalez-Ballestero2019}, obtaining the following analytical expression:
\begin{equation}\label{Shet}\begin{aligned}
&S_{\mathrm{c}}(\omega)=\\
&\frac{
\begin{aligned}
16\bigg[ & 4 g_x^4 \kappa \Omega_x^2\left(\omega^2-\Omega_y^2\right)^2\\+&g_x^2 \Omega_x\left(\omega^2-\Omega_y^2\right)\big[\Gamma_x \Omega_x\left(\omega^2-\Omega_y^2\right)\left(4(\Delta-\omega)^2+\kappa^2\right)\\
+&8 g_y^2 \kappa \Omega_y\left(\omega^2-\Omega_x^2\right)\big]\\
+&g_y^2 \Omega_y^2\left(\omega^2-\Omega_x^2\right)^2\left(\Gamma_y\left(4(\Delta-\omega)^2+\kappa^2\right)+4 g_y^2 \kappa\right)\bigg]
\end{aligned}}{\begin{aligned}\pi\bigg[\big[ &4 \Delta^2\left(\omega^2-\Omega_x^2\right)\left(\omega^2-\Omega_y^2\right)\\
+&16 \Delta\left(g_x^2 \Omega_x\left(\omega^2-\Omega_y^2\right)+g_y^2 \Omega_y\left(\omega^2-\Omega_x^2\right)\right)\\
+&\left(\kappa^2-4 \omega^2\right)\left(\omega^2-\Omega_x^2\right)\left(\omega^2-\Omega_y^2\right)\big]^2\\
+&16 \kappa^2 \omega^2\left(\omega^2-\Omega_x^2\right)^2\left(\omega^2-\Omega_y^2\right)^2\bigg].\end{aligned}}
\end{aligned}\end{equation}
To extract from our data the experimental values for $g_{x,y}$ and $\Gamma_{x,y}$ we fit Eq.~\eqref{eq:heterodyne} to our time traces while setting $\kappa$ and $\Delta$ to the values from our calibration.
\\

\centerline{\textbf{Calculated heating rates}}

Using equations~\eqref{gammagas} and~\ref{gammarecoil} we calculate the heating rates of $x$ and $y$ motion using the parameters given in the main text and $T = \SI{300}{\kelvin}$ $n_{\rm r} = 1.439$, $w_x = \SI{1.023}{\micro\meter}$, $w_y = \SI{0.856}{\micro\meter}$, $\gamma = 4812\times p_{\rm gas, mbar}$ with $p_{\rm gas, mbar}$ the gas pressure in mbar. 
At $p_{\rm gas, mbar}=5\times 10^{-9}$ we obtain $\Gamma^{(g)}_x = 2\pi\times \SI{0.053}{\kilo\hertz}$, $\Gamma^{(g)}_y = 2\pi\times \SI{0.045}{\kilo\hertz}$, $\Gamma^{(r)}_x = 2\pi\times \SI{0.825}{\kilo\hertz}$ and $\Gamma^{(r)}_y =2\pi\times \SI{1.379}{\kilo\hertz}$. 
The total heating rate is thus dominated by photon recoil, $\Gamma^{(g)}_{x,y} \ll \Gamma^{(r)}_{x,y}$ and $\Gamma^{(r)}_{x,y}$ agrees with the values extracted from our measured PSDs.
We heavily suppress heating due to phase noise, as we position the particle in the cavity node and use a ultra-low phase noise laser source~\cite{Meyer2019}. Note, that the stated pressure reading of $p_{\rm gas} = \SI{5e-9}{\milli\bar}$ is measured close to the ion-getter pump. The actual pressure at the position of the nanoparticle could be larger. $\Gamma^{(g)}_{x,y}$ becomes comparable to $\Gamma^{(r)}_{x,y}$ at $p_{\rm gas, mbar} \approx 10^{-7}$, suggesting the gas pressure near the nanoparticle is lower than this value. 
%, $R=71.5$nm, $m=3.37\times 10^{-15}$g, $P=1.2$W
\\

\centerline{\textbf{Error in sideband thermometry}}

The effects of optical-mechanical mode hybridisation and the decoupling of the dark mode modify the spectra measured through the cavity. 
As conventional sideband thermometry does not take these effects into account when estimating occupation numbers, the method is affected by a systematic error.
To obtain this error, we theoretically calculate the occupation $\bar{n}_j$ estimated via sideband thermometry by numerically computing the maxima of Eq.~\eqref{Shet} and compensating for the cavity asymmetry as detailed above. 
We then compare this result with the exact phonon occupations $\bar{n}^{\text{model}}_j = \langle\hat{b}_j^\dagger \hat{b}_j\rangle$ within our approximations (negligible friction, $z$ motion uncoupled). 
For the results in Fig. 4(d-e) of the main text we fix $\Delta = 2\pi\times 240$kHz, $g_x = g_y = g$, $\Omega_x= \Delta-\delta$ and $\Omega_y=\Delta+\delta$, and sweep over the parameters $\delta$ and $g$. 
At each point (i.e. for each value of $\Omega_{x,y}$) the corresponding heating rates $\Gamma_{j}$ are calculated from Eqs.~\eqref{gammagas} and \eqref{gammarecoil}.\\
 
\newpage

\bibliography{Supplementary}